\documentclass[prx,twocolumn,showpacs,preprintnumbers,amsmath,amssymb,nofootinbib]{revtex4-2}

\usepackage{graphicx}
\usepackage{dcolumn}
\usepackage{xspace}
\usepackage[utf8]{inputenc}
\usepackage{float}
\usepackage{subfigure}
\usepackage{url}

\usepackage{color}

\newcommand{\eq}[1]{Eq.~(\ref{#1})}

\newcommand{\fig}[1]{Fig.~\ref{#1}}

\newcommand{\ev}{\,\mbox{eV}}
\newcommand{\kev}{\,\mbox{keV}}
\newcommand{\mev}{\,\mbox{MeV}}
\newcommand{\gev}{\,\mbox{GeV}}
\newcommand{\tev}{\,\mbox{TeV}}

\newcommand{\no}{\nonumber}
\newcommand{\nn}{\no\\}
\newcommand{\lt}{\left}
\newcommand{\rt}{\right}
\newcommand{\tr}{\mbox{Tr}\,}

\definecolor{BlueViolet}{rgb}{0.2, 0.00, 0.7}
\definecolor{Blue}{rgb}{0.15, 0.00, 0.9}
\definecolor{halayaube}{rgb}{0.4, 0.22, 0.33}
\definecolor{sanddune}{rgb}{0.59, 0.44, 0.09}
\usepackage[colorlinks=true,linkcolor=cyan,citecolor=blue,urlcolor=BlueViolet,hyperfootnotes=false]{hyperref}

\begin{document}

\preprint{IFIC/20-23}
\preprint{TTP20-023}
\preprint{P3H-20-022}
\title{\boldmath Modified majoron model for cosmological anomalies}

\author{Gabriela Barenboim}
\email{gabriela.barenboim@uv.es}
\affiliation{\normalsize \it 
Departament de Fisica Te\`orica and IFIC, Universitat de Val\`encia,  Avda. Dr. Moliner, 50,
E-46100 Burjassot, Val\`encia,
Spain}

\author{Ulrich Nierste}
\email{ulrich.nierste@kit.edu}
\affiliation{\normalsize \it 
 Institut f\"ur Theoretische Teilchenphysik (TTP),
  Karlsruher Institut f\"ur Technologie (KIT), 76131 Karlsruhe, Germany.}

\begin{abstract}
  The vacuum expectation value $v_s$ of a Higgs triplet field $\Delta$
  carrying two units of lepton number $L$ induces neutrino masses
  $\propto v_s$. The neutral component of $\Delta$ gives rise to two
  Higgs particles, a pseudoscalar $A$ and a scalar $S$.
The most general renormalizable Higgs potential $V$ for $\Delta $ and
the Standard-Model Higgs doublet $\Phi$ does not permit the possibility
that the mass of either $A$ or $S$ is small, of order $v_s$, while the
other mass is heavy enough to forbid the decay $Z\to A S$ to comply with 
LEP 1 data. We present a model with additional dimension-6 terms
in $V$, in which this feature is absent and either $A$ or $S$ can be
chosen light. Subsequently we propose the model as a remedy  to
cosmological anomalies, namely the tension between observed and predicted
tensor-to-scalar mode ratios in the cosmic microwave background and the
different values of the Hubble constant measured at different
cosmological scales. Furthermore, if  $\Delta$ dominantly couples to the
third-generation doublet $L_\tau=(\nu_\tau,\tau)$,  the deficit of
$\nu_\tau$ events at IceCube can be explained. The singly and doubly
charged triplet Higgs bosons are lighter than 280\gev\ and 400\gev,
respectively, and could be found at the LHC.
\end{abstract}
\maketitle
\renewcommand{\thefootnote}{\#\arabic{footnote}}
\setcounter{footnote}{0}

\section{\label{sec:intro}Introduction}
Although the Hot Big Bang Model and General Relativity are arguably very
robust, they work only if an additional piece is added to the game:
inflation. In order to explain the flatness, homogeneity and isotropy of
the Universe and the abscence of monopoles and other relics, a period of
inflation is crucial.

But inflation is a framework comprising 
countless different inflationary models. 
Clearly all of them
produce a flat, isotropic, homogeneous and relic-free Universe but each
one leaves some specific imprints (as the inhomogeneities pattern of the model at
hand in the CMB and structure formation) that can help us find out,
which one of the plethora of models in the market is the correct one.

During inflation two types of perturbations are produced: scalar or
matter perturbations and tensor (metric) perturbations (gravity
waves). Each one can be characterized by its amplitude and the
dependence on the scale of such amplitude. 
However, only a subset of two of these four quantities is
  independent and therefore all our insight of inflation is reduced to
two parameters generally chosen to be the spectral index $n_s$, {\it
  i.e.} the dependence on the scale of the matter perturbations, and the
tensor to scalar (amplitude) ratio $r$.  This is the reason why all the
inflationary models reduce to lines, points or regions in the $n_s-r$
plane.

As a consequence to discriminate which region is favored by experiments
is also to select which inflationary models remain in the game. The
theoretical guidance at this stage is crucial. Specific particle physics
models with their matter content and interactions should help shed some
light on which are the inflationary potentials worth considering, while
at the same time making predictions which can be tested elsewhere.

In recent years tensions of cosmological data with the predictions
  based on the SM and the $\Lambda$CDM model have emerged and a light
  scalar boson $\phi$ interacting with neutrinos has been considered to
  alleviate these tensions.  Specifically, 
favored regions for  the ratio $r$ of tensor (metric) to scalar
(matter) perturbations inferred from the anisotropies of the cosmic
microwave background (CMB) and the spectral index $n_s$ can be
significantly modified and therefore the selection rule for successful
inflationary models is vastly affected \cite{Oldengott:2017fhy,
  Lancaster:2017ksf, Kreisch:2019yzn, Barenboim:2019tux}. Furthermore,
the Hubble constant determined from local measurements disagrees with
the value inferred from CMB data and new neutrino interactions might
  remedy this as well \cite{Blinov:2019gcj,Escudero:2019gvw}.

Historically, the first interest into light scalars interacting with
  neutrinos was driven by the attempt to build \emph{majoron}\ models
  breaking lepton number spontaneously, {first realized through an SU(2) singlet
  \cite{Chikashige:1980qk,Chikashige:1980ui} or triplet
  \cite{Gelmini:1980re} field, or with several fields in both
  representations \cite{Schechter:1981cv}.   
  The triplet}
  models (and those employing doublet fields) did not comply with LEP 1
  data on invisible $Z$ decays \cite{Berezhiani:1992cd} {shifting}
  the focus {entirely to}
  SU(2) singlet fields $\phi$ 
  \cite{Berezhiani:1992cd,Burgess:1993xh,Bamert:1994hb}. {But}
  couplings of singlets 
  to active neutrinos
  are tiny,  because SU(2) forbids such couplings at dimension-4 level
    and the effective coupling is necessarily suppressed by small
    mixing angles (e.g.\ between active and sterile neutrinos). E.g.\
    Ref.~\cite{Berezhiani:1992cd} finds coupling constants below
    $10^{-3}$. As we will see below, such small couplings do not permit
    solutions to the cosmological problems in the most interesting
    region with $m_\phi>30\kev$.
   Thus to date there is no viable
   model supporting the idea of Refs.~\cite{Oldengott:2017fhy,
  Lancaster:2017ksf, Kreisch:2019yzn, Barenboim:2019tux,
  Blinov:2019gcj,Escudero:2019gvw} {in this mass range.}

In the following we will present a model of neutrinos that can not only
modify the allowed region in the $n_s-r$ plane changing this way the
inflationary models that survive the experimental scrutiny but also
provides a {viable modification} of the {SU(2) triplet} majoron
idea \cite{Gelmini:1980re}. {Since triplet fields have
  renormalizable, dimension-4 couplings to leptons, even ${\cal O}(1)$
  couplings of the majoron to active neutrinos are possible a
  priori. Thus we reopen a large portion of the parameter space with
  consequences for other applications of a light scalar coupling to
  neutrinos. Furthermore, our model makes predictions beyond cosmology
  and neutrino physics and}\ can be tested in forthcoming experiments.

This paper is organized as follows: In 
the following section we present a
class of majoron models in which either $S$ or $A$ is light, while the
other boson is heavy enough to forbid  $Z\to AS $. 
Next we discuss  phenomenological consequences and ``smoking gun'' features.
Finally we conclude.

\section{\label{sec:model}The model}

An SU(2) triplet Higgs field
\begin{align}
   \Delta = \begin{pmatrix}  
       \frac{\delta^+}{\sqrt2} & \delta^{++} \\  
       v_s + \frac{h_s+i a_s}{\sqrt2} &     - \frac{\delta^+}{\sqrt2}  
      \end{pmatrix}  \label{defd}
\end{align} 
developing a vacuum expectation value (vev) $v_s$ in the neutral
component generates Majorana masses of light neutrinos in a natural way
via its coupling to the lepton doublets. Electroweak precision data
imply $v_s\ll v$, where $v=174\gev$ is the vev of the doublet Higgs
field $\Phi$ of the Standard Model (SM). The physical Higgs fields are
mixtures of the components of $\Delta$ and  $\Phi$; in particular there
are two extra neutral Higgs bosons, a scalar $S$ and a pseudoscalar {$A$}.
These approximately coincide with $h_s$ and $a_s$, respectively. 
By assigning two units of lepton number $L$ to $\Delta$ and choosing
an U(1)$_L$-invariant Higgs potential one arrives at a model which breaks
U(1)$_L$ spontaneously. Then $A$ is a massless Goldstone mode, the
\emph{majoron} \cite{Chikashige:1980qk,Chikashige:1980ui,Gelmini:1980re}. 

Postulating
an effective interaction of neutrinos with a light scalar $\phi$,
\begin{align}
  L_{\rm eff}= g_{\alpha\beta} \bar\nu_\alpha \nu_\beta \phi,  \label{leff}
\end{align}
where $\nu_{\alpha}$ generically denotes any of the light left-handed
neutrino fields or its charge-conjugate, one can alleviate the
  tensions in cosmological data mentioned in the introduction while
simultaneously modifying the $n_s-r$ region to include well motivated
inflationary models which were previosly ruled out \cite{Oldengott:2017fhy,
  Lancaster:2017ksf, Kreisch:2019yzn, Barenboim:2019tux,
  Blinov:2019gcj}. In this paper we only consider couplings of $\phi$
to $\tau$-neutrinos $\nu_\tau$, for which terrestrial and astrophysical
data do not imply strict bounds and it is not relevant for us whether
$L$ is identified with the total lepton number or with $L_\tau$.
{We aim at the formulation of a minimal triplet majoron model addressing the
  cosmological anomalies and the majoron couplings to $\nu_e$ and $\nu_\mu$
  are neither conceptually nor phenomenologically relevant for this.}

$L_{\rm eff}$ is not gauge invariant, a meaningful interaction must be
formulated in terms of lepton doublets. Identifying $\phi$ with the
majoron amounts to completing $L_{\rm eff}$ to
\begin{align}
 \! L_y^\Delta =\; &  \frac{y^\Delta_\tau}{2} \overline{L_3^c} \Delta L_3 + h.c. \nn  
 \supset \; &
           -   \frac{m_{\nu_\tau}^\Delta}{2} (\bar\nu_\tau
       \nu_\tau^c+ \overline{\nu_\tau^c} \nu_\tau )   \nn
       & 
       - \; \frac{y^\Delta_\tau}{{2\sqrt2}}\,   \big[ (h_s+i a_s)   {\overline{\nu_\tau^c} \nu_\tau } 
   \; +\;                                                                                     
              (h_s - i a_s)  {\bar\nu_\tau \nu_\tau^c}) \big],
\label{yuk}
\end{align} 
where $L_3= (\nu_\tau ,\tau)^T$, $L_3^c= (\tau^c , -\nu_\tau^c)^T$, and
$ m_{\nu_\tau}^\Delta= y^\Delta_\tau v_s $ is the contribution of
$\Delta$ to the Majorana mass of $\nu_\tau$. However, all known Higgs
potentials predict that $S$ and $A$ are either both light, with masses
around $v_s$, or both heavy, with masses of order $v$ or larger. This
finding holds for the most general renormalizable Higgs potential,
irrespective of whether $L$ is broken spontaneously or explicitly. For
this reason the original {triplet} majoron models were discarded with
the advent of LEP 1 data on the invisible $Z$ width which left no room
for the decay $Z\to AS $, whose decay rate is entirely fixed by the
value of the SU(2) gauge coupling.\footnote{The decays $Z\to SS$ and
  $Z\to AA$ are forbidden by CP invariance of the electroweak gauge
  interaction.}  Finally, in models in which $\phi$ in \eq{leff} is a
singlet field mixing with a heavy $S$ or $A$ the coupling
$g\equiv g_{\tau\tau}$ is suppressed by a tiny mixing angle and is far
too small to solve the cosmological tensions {in the interesting region with
  $m_\phi>30\kev$}. The same remark applies, if a singlet $\phi$ couples
to heavy sterile neutrinos which mix with $\nu_\tau$.

For the phenomenological analysis {of the relevant collider bounds
  and astrophysical constraints discussed below} it does not matter whether $S$ or $A$
is the light scalar corresponding to $\phi$ in  \eq{leff} and for
definiteness we consider the case $\phi=S\simeq h_s$. Then
\eq{yuk} entails $g=- y^\Delta_\tau/{(2\sqrt{2})}$.  

Global fits to cosmological data constrain the combination \cite{Barenboim:2019tux}
\begin{align}
  G_{\rm eff} =&\; \frac{g^2}{m_\phi^2} \;=\; \frac{y^{\Delta\, 2}_\tau}{{8} \,m_S^2},  
\label{geff}
\end{align}  
while successful Big Bang Nucleosythesis (BBN) is sensitive to the mass
and coupling of the scalar particle in a different combination. More
specifically, BBN is very sensitive to the amount of extra radiation.
The observed primordial abundances of deuterium and helium set strong
constraints on the coupling and the mass of our scalar particle. These
bounds are reflected in figure 1, where it can be seen that two
  regions are consistent with BBN and at the same time result in a
  $G_{\rm eff}$ able to change the CMB temperature and polarization
  spectra.\footnote{{As was shown in \cite{Barenboim:2019tux} 
 $G_{\rm eff}$  in the ballpark of 3-5 $ \cdot 10^{-2}$ MeV$^{-2}$ 
 opens up the allowed window in the $n_S-r$ plane,  significantly affecting
 the selection criteria for acceptable models of inflation.}}
 A \emph{heavy} or \mev\ region with
  ${0.03}\mev \leq m_S \leq 1 \mev$ and a \emph{light} or \kev\ region
  with $m_S \leq 0.1$ \kev.
\begin{figure}[t]
\centering
\includegraphics[width=0.4\textwidth]{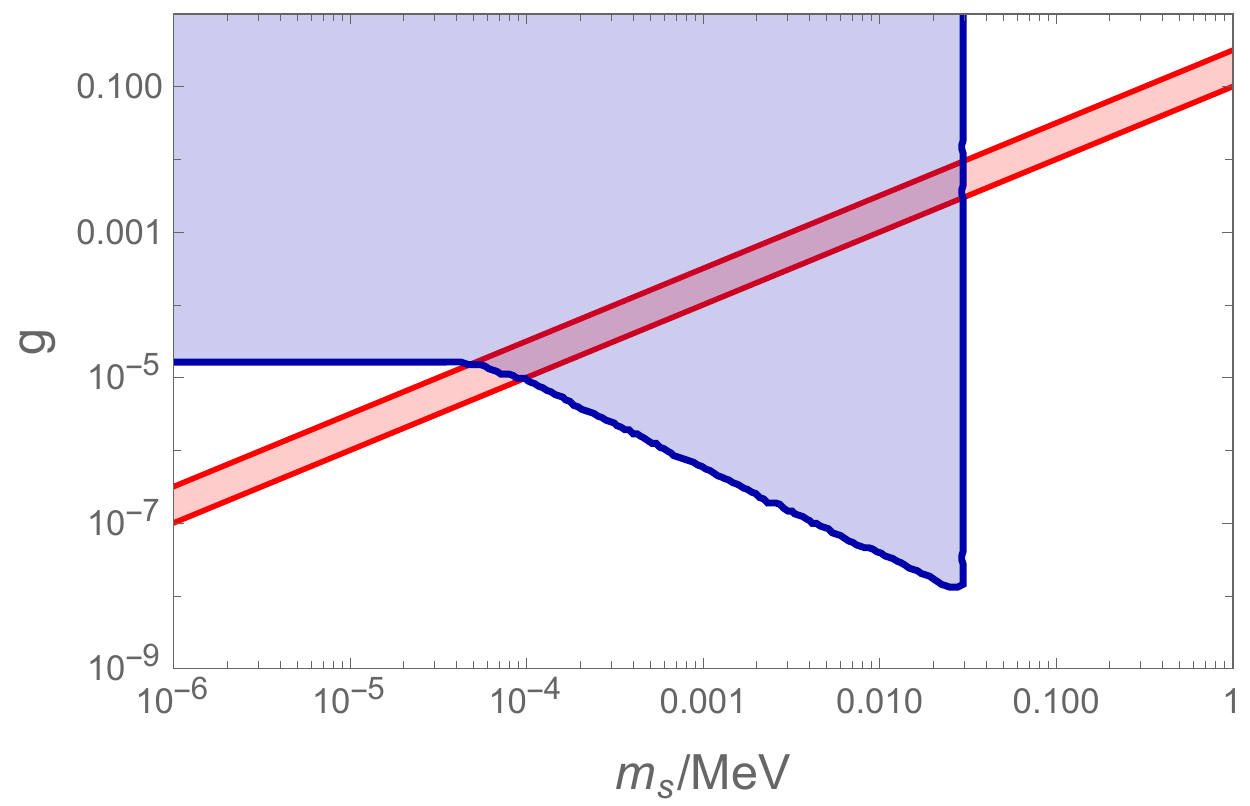}
\caption{The contour of the extra radiation
  $\Delta N_{{\rm eff}}\equiv N_{{\rm eff}}- 3= 0.6 $ at a temperature
  of 1 \mev\ in the $m_S$-$g$ plane. The (light-blue) region above the
  blue line is forbidden by primordial helium and deuterium
    abundances, thus $m_S$ is in the sub-\kev\ range or
  $m_S\geq 30\kev$.  
  The red region corresponds to
  $10^{-2}/\mev^2 <G_{\rm eff} <10^{-1}/\mev^2$, which is the 2$\sigma$
  CMB favored region for $G_{\rm eff} >10^{-4}/\mev^2$
  \cite{Barenboim:2019tux}. Unlike SU(2) singlet models our triplet model
  covers the whole allowed parameter region, up to $m_s\sim 1 \mev$.
  {For the displayed region with $g\geq 10^{-7}$ the bound
    $m_{\nu_\tau}\leq 1\,$eV implies $v_s\leq 3.5\,$MeV.}
   \label{fig:bbn}} ~\\[-3mm]
\hrule
\end{figure}

Bounds from meson decays are irrelevant in our case (and therefore not
shown) as our new interaction concerns only tau neutrinos. We further
stress that none of the bounds derived on $L_{\rm eff}$ in \eq{leff} from
$\tau$ or tauonic $Z$ decays (see e.g.\ \cite{Brdar:2020nbj} for a
recent study) applies to a complete model like ours: The proper
cancellation of infrared singularities (for $m_\phi\to 0$) requires the
inclusion of virtual corrections, which in turn involve also heavy
fields (like the charged components of $\Delta$) to cancel ultraviolet
divergences.

The Higgs potential of $\Delta$ in \eq{defd},
$\Phi =  (\phi^+ , v + \frac{h+i  a}{\sqrt2})^T$, and 
$ \Phi^c = (v + \frac{h-i  a}{\sqrt2},  -\phi^- ) $ reads:
\begin{align} 
  V =&\;    -\mu^2\Phi^\dagger \Phi \;-\; \mu_\Delta^2  \tr(\Delta^\dagger
       \Delta) \; +\; 
       \lambda (\Phi^\dagger \Phi )^2 \nn
       &\; +\; 
        \lambda_\Delta \lt[ \tr(\Delta^\dagger \Delta)\rt]^2 \;  +\;
        \alpha_1 \, \Phi^\dagger \Delta^\dagger \Delta \Phi \; +\;
         \alpha_2 \, \Phi^\dagger \Delta \Delta^\dagger \Phi \nn
         & \;+\;
           \alpha_3 \,\Phi^\dagger \Phi \, \tr(\Delta^\dagger \Delta)
           \;   - \; \beta \,
   \lt(\Phi^{c\,\dagger} \Delta^\dagger \Phi  + \Phi^\dagger
        \Delta \Phi^c \rt)   \nn 
   &
     \;  + \;
      \delta_1\, 
       \lt( \Phi^{c\,\dagger} \Delta^\dagger \Phi +
                       \Phi^\dagger \Delta \Phi^c \rt)^2    \nn 
 &     \;  - \;
     \delta_2 \, \lt( \Phi^{c\,\dagger} \Delta^\dagger \Phi - 
                       \Phi^\dagger \Delta \Phi^c \rt)^2   
  \label{pot} .
\end{align}
$V$ is complete up to terms of dimension 4\footnote{The term
  $\lambda_\Delta^\prime \tr(\Delta^\dagger \Delta^\dagger ) \tr(\Delta
  \Delta)$ is phenomenologically irrelevant.} and the dimension-6 terms
involving $\delta_{1,2}$ are instrumental to lift the mass of either $S$
or $A$ above $M_Z$. All parameters are chosen real, so that $V$ is
invariant under charge conjugation $C$, and we only consider solutions
with real vevs.  $L$ is a good symmetry of $V$ for
$\beta=\delta_1-\delta_2=0$. {Any other dim-6 operator can be
  expressed as a linear combination of the terms in \eq{pot} and a
  $\Delta L=0$ operator, which does not contribute to the $A$-$S$ mass
splitting.}
The minimization conditions
$\partial V/\partial v=0=\partial V/\partial v_s$ read:
\begin{align} 
  \mu^2 = &\; 2 \lambda v^2 \,+\,{\cal O} \left( \frac{v_s^4}{v^2}\rt), \nn
    \beta = &\; v_s \lt( \frac{m^2}{v^2} +4 \delta_1 v^2 + 
              2 \lambda_\Delta \frac{v_s^2}{v^2} \rt) 
                 \label{min} \\
  \mbox{with }\qquad
  m^2 \equiv &\;   -   \mu_\Delta^2 + \alpha_2 v^2 + \alpha_3 v^2 \label{defm2}
\end{align}
The parameter $m^2$ will govern the mass of the desired light state.
Avoiding fine-tuning between different terms in \eq{defm2} we must have 
\begin{align} 
 \mu_\Delta^2 ,  \alpha_{2}v^2,  \alpha_{3}v^2  & \leq {\cal O}(m^2), &\qquad 
                                                  \label{hier}
\end{align}
Using the minimization conditions in \eq{min}
we trade $\mu$ and $\beta$ for $v$ and $v_s$ in the formulae below for fields
and  masses.
Neglecting ${\cal O} \lt(\frac{v_s^2}{v^2}\rt) $ terms the
physical states are
\begin{align}
  G &= a +  2 \frac{v_s}{v} \, a_s , \qquad\quad
    A\;=\; a_s - 2 \frac{v_s}{v} \, a, \label{psc} \\
H & = h - \, \frac{2 \mu_\Delta^2+8\delta_1 v^4 }{m^2-
                        4  \lambda v^2 +4\delta_1 v^4} \, \frac{v_s}{v}
    \, h_s, \nn 
   S & = h_s + \, \frac{2 \mu_\Delta^2+8\delta_1 v^4 }{m^2-
                        4  \lambda v^2 +4\delta_1 v^4}\, \frac{v_s}{v}  \, h, \label{sc} \\
  G^+ & = \phi^+ + \sqrt2 \frac{v_s}{v} \,  \delta^+ , \quad
  H^+ \; =\;  \delta^+ - \sqrt2 \frac{v_s}{v} \,  \phi^+ ,  \\
H^{++} & = \delta^{++}.
\end{align}
Here $G$ and $G^+$ are the massless Goldstone bosons eaten by $Z$ and
$W^+$, respectively.
Neglecting subdominant terms the desired (squared) Higgs masses read
\begin{align}
  m_A^2 &= 4 \delta_2 v^4 + m^2 + 2  \lambda_{\Delta} v_s^2
                                                                                  \label{lightmasses} \\
 m_S^2  &=  4 \delta_1 v^4 +  m^2 + 6 \lambda_\Delta v_s^2
                                                                        \label{lightmasses2} \\
  m_H^2&= 4 \lambda v^2 =(125\gev)^2, \label{heavymasses}\\
   m_{H^{++}}^2 &=2  m_{H^+}^2= \alpha_1 v^2 .  \label{heavymasses2}
\end{align}

We start our discussion with the role of U(1)$_{L}$  symmetry
in  $V$. For this it is helpful to
use \eq{min} to trade $m^2+ 2 \lambda_{\Delta} v_s^2$ for $\beta$ in
$m_A^2 $ in \eq{lightmasses} to find
\begin{align}
 m_A^2 &= 4 (\delta_2 - \delta_1) v^4 + \frac{\beta v^2}{v_s} .\label{masym} 
\end{align}
For exact U(1)$_{L}$ symmetry with $\beta=0$ and
$\delta_1=\delta_2$ we verify that $A$ is the massless
Goldstone boson of this broken symmetry. This solution corresponds to the
vanishing of the bracket in \eq{min}. (For the second solution  with
$v_s=0$ the symmetry is unbroken and \eq{masym} does not hold.)
An interesting case is $\beta=\delta_1=0$ with $\delta_2\neq 0$:  
U(1)$_{L}$ is explicitly broken by a higher-dimensional term
only: The minimization equation (\ref{min}) is not sensitive to this
term and features spontaneous U(1)$_{L}$ as in the original,
renormalizable majoron models. The phenomenological effect of $\delta_2\neq 0$ is 
to render $m_A$ massive, with the possibility of $m_A> M_Z$,
and we may view this case as \emph{spontaneous symmetry breaking without
  Goldstone}. 

For fixed $G_{\rm eff}$ the perturbativity limit $y^{\Delta}_\tau
\lesssim 2$ entails an upper limit on $m_S$ through \eq{geff}. In the
scenario with spontaneous U(1)$_{L}$ breaking (where $\beta=0$)
one necessarily has
$m\lesssim v_s$. As a consequence, $ m_{\nu_\tau}^\Delta= y^\Delta_\tau v_s
\lesssim 1\ev$ pushes   $y^{\Delta}_\tau$ and $m_S$ far below
their otherwise theoretically allowed upper bounds. 
Specifically,
$m_S \lesssim 10 \kev$ for $G_{\rm eff}\geq 10^{-4}\mev{}^{-2}$ and the BBN
constraint of figure 1 tightens this to $m_S \lesssim 0.3 \kev$.
In the scenario,
with explicit  U(1)$_{L}$ breaking, however,
one easily infers from \eq{min} that one can choose $m$
(and thereby $m_S$) and $v_s$ independently thanks to the free parameter 
$\beta$.

It is easy to find a UV completion generating the dim-6 terms in
\eq{pot}: Consider heavy real scalar singlets $\chi_{1,2}$ coupling as
\begin{align}
  V_s = (\rho_1+ i\sigma_1) \chi_1 \Phi^{c\,\dagger} \Delta^\dagger \Phi + 
  (\rho_2+ i\sigma_2) \chi_2 \Phi^{c\,\dagger} \Delta^\dagger \Phi + \mbox{H.c.}
\label{eq:uv}
\end{align}
with real $\rho_{1,2}$, $\sigma_{1,2}$.
\begin{figure}[t]
\centering
\includegraphics[width=0.4\textwidth]{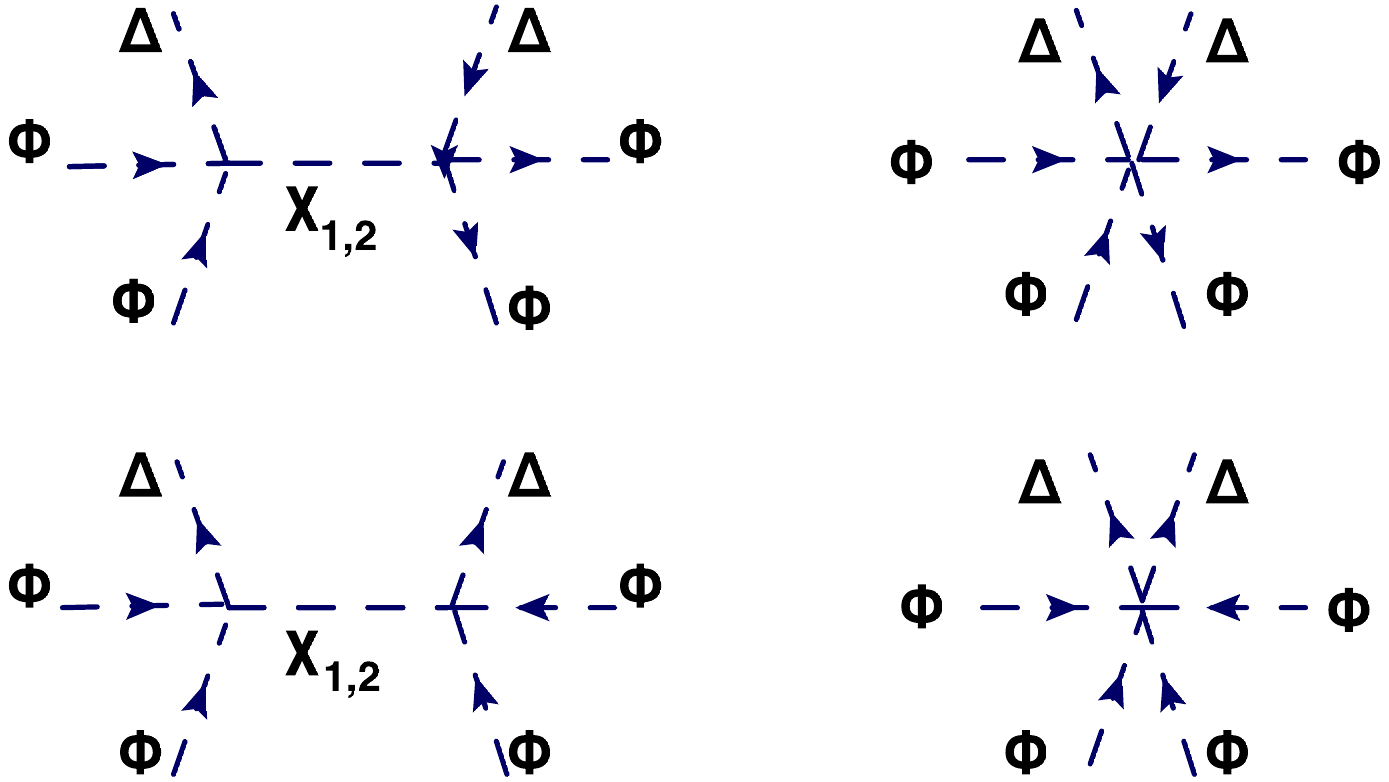}
\caption{{UV completion with heavy real scalars $\chi_{1,2}$: $V_s$ in \eq{eq:uv} contains
  the renormalizable couplings entering the left diagrams. Integrating
  out $\chi_{1,2}$ in these diagrams generates the depicted dim-6 operators
  $\Phi^{c\,\dagger} \Delta^\dagger \Phi \, 
  \Phi^\dagger \Delta \Phi^c $ (top)   
  and  
  $\lt(\Phi^{c\,\dagger} \Delta^\dagger \Phi \rt)^2$ (bottom). By flipping all arrows in the
  bottom row one generates the hermitian conjugate of the second operator.
   \label{fig:uv}}} ~\\[-3mm]
\hrule
\end{figure}
Under charge conjugation we
have $\Delta \leftrightarrow \Delta^*$, $\Phi\leftrightarrow \Phi^*$.
Choosing further $\chi_1 \leftrightarrow \chi_1$,
$\chi_2 \leftrightarrow - \chi_2$ {under $C$}
and demanding $C$ invariance of $V_s$ implies
$\sigma_1=\rho_2=0$. Integrating out
the heavy singlet fields {as shown in \fig{fig:uv}}
gives $\delta_1=\rho_1^2/({2}m_{\chi_1}^2)$ and
$\delta_2=\sigma_2^2/({2}m_{\chi_2}^2)$, and e.g.\
$m_{\chi_1} \gg m_{\chi_2}$ produces the scenario with heavy $A$ and
light $S$. {Thus the UV completion in \eq{eq:uv} generates the desired
  dim-6 terms in \eq{eq:uv} without producing any other dim-6
  operators.} 

Instead of invoking $C$ symmetry one can also work with
$L$, by assigning $L={2}$ to $\chi= \chi_1 + i
\chi_2$ enforcing $\rho_2+ i\sigma_2=i(\rho_1+
i\sigma_1)$. {Compared to the minimal model described above} the UV
sector must be {richer} and break
$L_{(\tau)}$ in a way to produce the desired mass hierarchy. Loop
effects and/or a small vev of $\chi_1$ may render $\beta\neq
0$, {which is a welcome feature as discussed in the paragraph above
  \eq{eq:uv}.  {Note that
    $\chi$ does not need to acquire a vev. If one chooses to consider
    scenarios with a non-zero
    $\chi$ vev, this vev must be small to avoid large corrections to
    $\beta$ spoiling $|\beta| \lesssim {\cal O}
    (v_s)$ implied by \eq{min}. Note that all other dim$\,\lesssim 4$
    terms in $V$ conserve $L$ and therefore do not receive contributions
    linear in the $L$-breaking vev of $\chi$.}  We do not aim at an exhaustive
    discussion of all attractive UV completions here. Instead, we
  consider it as an advantage that
  $V$ in \eq{pot} allows us to fully study the low-energy phenomenology
  (including loop effects where needed) without specifying the
  {underlying fundamental theory}.

\section{Phenomenology\label{sec:pheno}}
Studies of perturbativity for the SM \cite{Nierste:1995zx} and 2HDM
\cite{Cacchio:2016qyh,Murphy:2017ojk} have shown that self-couplings
should be smaller than $\approx 5$.  Applying this bound to $\alpha_1$
in \eq{heavymasses2} implies that $m_{H^{++}}\lesssim 400\gev$ and
$m_{H^+}\lesssim 280\gev$.  Current collider bounds are much weaker,
because the production of these heavy charged Higgs bosons is an
electroweak gauge process (e.g.\ vector boson fusion at the LHC).
Favorable decays are $H^{++}\to \tau^+\tau^+$ and
$H^{+}\to \tau^+\nu_{\tau}$, unless $y^\Delta_\tau$ is too small. In the
latter case one must resort to gauge-coupling driven decays like
$H^{+}\to W^+ S, W^+ A$.  For a cut-off scale of $\Lambda \sim 0.5 \tev$
(and ${\cal O}(1)$ couplings in the UV completion) we have
$\delta_{2} v^2 \sim 0.1$ and \eq{lightmasses2} gives
$m_A\sim 120\gev$. $A$ is produced through gauge interactions,
and now a small $y^\Delta_\tau$ is welcome to suppress the decay into
neutrinos. Detection through $A\to Z S$ will fail if $M_A-M_Z$ is
smaller than the trigger threshold for missing transverse momentum.
It is therefore advisable to focus on the searches for the charged bosons.   

The model can be tested by its astrophysical signatures as
well. Depending whether our scalar field is in the \mev\ or \kev\ range
different signals can be expected.  As mentioned before, CMB expriments
are  sensitive only to $G_{\rm eff}$ and therefore both ranges give
exactly the same phenomenology cosmology-wise. This is not the case
regarding astrophysical experiments. For scalars in the \mev\ range, the
interaction introduced will make high energy ($\sim$TeV) tau neutrinos from
astrophysical sources scatter resonantly with the CMB tau neutrinos
and therefore a deficit of tau neutrinos can be expected. More precisely
a dip in the tau neutrino spectrum corresponding to the resonant energy
\cite{Barenboim:2019tux}
\begin{equation}
E_{\rm resonant}\simeq \frac{m_\phi^2}{2 m_{\nu_\tau}}
\end{equation} 
is to be expected in experiments like IceCube and KM3Net\footnote{Note
  that when the tau neutrino mass drops below the neutrino CMB
  temperature the resonance energy becomes independent of the neutrino
  mass. In this case the resonance falls into the IceCube sensitivity range,
  unless the mass of the neutrino is close to the current cosmological
  limit}.
Remarkably, IceCube seems to be seeing a deficit in tau neutrinos
although the effect is not significant yet.

For scalars in the \kev\ range the resonant energies involved make it
ideal to detect such interactions in experiments sensitive to the
diffuse supernova neutrino background, like T2HK. A detailed analysis of
both signals will be given elsewhere.

{As mentioned before, another very attractive property of this model is that it provides a 
higher value of the Hubble constant $H_0$, {\it i.e.} at 95\% C.L. 
$ H_0 = 70.06^{+2.21}_{-2.31} \mathrm{km/s/Mpc}$ as compared with the 
 $\Lambda$CDM  value of $ H_0 = 68.35^{+1.94}_{-1.84}    \mathrm{km/s/Mpc}$ . 
 
This is specially  welcome, 
as the CMB provides a 2-3$\sigma$ \cite{Aghanim:2018eyx} lower value of the Hubble 
constant than local measurements do \cite{Riess:2016jrr}. In a simple 1-parameter 
extension, by means of $G_{\mathrm{eff}}$ this tension is only weakened but not 
resolved. However, including also $N_{\rm eff}$ and $\sum m_{\nu}$ as free 
parameters to the analysis completely relaxes  the tension in the Hubble parameter. 
Such an extension is however beyond the scope of the present work.
}

\section{Conclusions\label{sec:conc}}
The possibility of a light scalar particle interacting with neutrinos receives a lot of
attention to alleviate several tensions in cosmological data.
The particle physics community is interested in this kind of interaction
as a means to break lepton number $L$ spontaneously, providing a natural
framework for neutrino Majorana masses. So far all cosmological analyses
have employed the effective interaction of \eq{leff}, which violates
electroweak SU(2) symmetry. In this paper we have presented a viable
realization of a model of a scalar interacting with neutrinos, by
complementing the original SU(2) triplet models (featuring spontaneous
or explicit $L$ violation) with higher-dimensional terms and devising
possible UV completions generating these terms.  With our lagrangian
one can now  consistently calculate constraints from $Z$,
charged lepton, and  meson decays, which was not possible with
\eq{leff}. To bypass these constraints we have discussed the
cosmological and astrophysical implications for the case that the light
scalar couples dominantly to tau neutrinos. Depending on the mass range
of the scalar particle, characteristic  signals are possible at the
IceCube or T2HK experiments. The LHC can search for the charged
members of the SU(2) triplet, whose masses must be below 400\gev.
For appropriate choices of the parameters our Higgs potential
conserves $L$ at the level of the renormalizable terms and shares
essential features of the orginal {triplet} majoron model \cite{Gelmini:1980re}.

\vspace{6mm} {\it Acknowledgments.}~~~ 
GB acknowledges support from FPA2017-845438 and the Generalitat Valenciana under grant PRO\-ME\-TEOII/2017/033, also partial support from the European Union FP10 ITN ELUSIVES (H2020-MSCAITN-2015-674896) and INVISIBLES-PLUS (H2020- MSCARISE- 2015-690575).
The work of U.N. is
supported by project C3b of the
DFG-funded Collaborative Research Center TRR 257, ``Particle Physics
Phenomenology after the Higgs Discovery''.

\end{document}